\documentclass[aps,pre,floats,twocolumn,twoside,floatfix,superscriptaddress]{revtex4}

\usepackage{graphicx}
\usepackage{amsmath,amssymb}
\usepackage{psfrag}
\usepackage[utf8]{inputenc} 

\newcommand{\Tc}{T_{\scriptstyle \rm c}}
\newcommand{\Tf}{T_{\scriptstyle \rm f}}
\newcommand{\tp}{t_p} 
\newcommand{\tpeak}{t_{\scriptscriptstyle \rm peak}}

\newcommand{\Heq}{H_{\scriptstyle\rm eq}}

\newcommand{\taui}{\tau} 
\newcommand{\taup}{\tau_p} 
\newcommand{\zp}{z_p} 

\begin{document}

\title{Dynamical cluster size heterogeneity}

\author{Amanda de Azevedo-Lopes}
\affiliation{Instituto de F\'\i sica, Universidade Federal do
Rio Grande do Sul, CP 15051, 91501-970 Porto Alegre RS, Brazil}

\author{André R. de la Rocha}
\affiliation{Instituto de F\'\i sica, Universidade Federal do
Rio Grande do Sul, CP 15051, 91501-970 Porto Alegre RS, Brazil}

\author{Paulo Murilo C. de Oliveira}
\affiliation{Instituto de Física, Universidade Federal Fluminense, 
Av. Litorânea s/n, 24210-340 Boa Viagem, Niterói, RJ, Brazil}
\affiliation{Instituto Nacional de Ciência e Tecnologia - Sistemas
Complexos, Rio de Janeiro RJ, Brazil}

\author{Jeferson J. Arenzon}
\email{arenzon@if.ufrgs.br}
\affiliation{Instituto de F\'\i sica, Universidade Federal do
Rio Grande do Sul, CP 15051, 91501-970 Porto Alegre RS, Brazil}
\affiliation{Instituto Nacional de Ciência e Tecnologia - Sistemas
Complexos, Rio de Janeiro RJ, Brazil}

\date{\today}

\begin{abstract}
Only recently the essential role of the percolation critical point has been considered on the dynamical properties of connected regions of aligned spins (domains) after a sudden temperature quench. In equilibrium, it is possible to resolve the contribution to criticality by the thermal and percolative effects (on finite lattices, while in the thermodynamic limit they merge at a single critical temperature) by studying the cluster size heterogeneity, $\Heq(T)$, a measure of how different the domains are in size. We here extend this equilibrium measure and study its temporal evolution, $H(t)$, after driving the system out of equilibrium by a sudden quench in temperature.
 We show that this single parameter is able to detect and well separate the different time regimes, related to the two time scales in the problem, the short, percolative and the long, coarsening one.
\end{abstract}

\maketitle
\section{Introduction}

The ferromagnetic Ising model displays relatively homogeneous
configurations when equilibrated either at temperatures $T\ll T_c$ or $T\gg T_c$, where $T_c$ is its critical temperature. In the former case, thermal fluctuations in the giant, equilibrium background cluster of aligned spins are energetically inhibited but increase in probability with temperature. In the opposite limit, well above $T_c$, large domains of parallel spins are unstable against the thermal noise, which breaks them into small clusters whose average size decreases with temperature. At these extreme limits, the size diversity is smaller than that found close to $T_c$, where the distribution of allowed sizes is very broad, with a fully developed power-law (in the thermodynamic limit, $L\to\infty$).
Because neighboring parallel spins are not necessarily correlated, besides the geometric clusters described above,   Coniglio-Klein (CK) clusters~\cite{CoKl80} may be built by removing a temperature dependent fraction of the parallel pairs from the geometric clusters. These so called physical clusters have been useful in developing powerful simulation algorithms~\cite{SwWa87,Wolff89} and to unveil geometric properties for both models and experimental systems~\cite{Bray94,SiArDiBrCuMaAlPi08,CaCuGaVa14,DePr14,Timonin19,Almeida20} that characterize both the equilibrium critical behavior~\cite{CoKl80,CaZi03} and the out of equilibrium dynamics~\cite{Bray94}. 
The domain size distribution only becomes dense in the infinite size limit or after ensemble averages are taken, while for a single, finite sample, space constraints forbid the presence of every possible cluster size, the distribution gets truncated and sparse, subject to sample-to-sample fluctuations. A simple, global measure of the heterogeneity of a finite equilibrium configuration  was introduced~\cite{LeKiPa11,NoLePa11,LvYaDe12,JoYiBaKi12,RoOlAr15}, only taking into account whether a given size is present in a configuration. The cluster size heterogeneity ($H$) is defined as the number of distinct cluster sizes, irrespective of the number of equally sized domains, that are present in a finite-size sample. 


The results for the equilibrium cluster size heterogeneity $\Heq(T)$ of the geometrical domains in the 2d Ising model show
a double peak structure at two very distinct temperatures.
The small peak at $T_1\simeq T_c$, associated with the thermal
transition~\cite{JoYiBaKi12}, is only observed for sufficiently large
systems~\cite{RoOlAr15}. The peak grows as $\Heq(T_1)\sim L^{d/\taui}$ where
$\taui=379/187\simeq 2.027$ is the Fisher exponent associated with the power-law
cluster size distribution at the critical temperature of the Ising model~\cite{StVa89,JaSc05a}.
In spite of the thermal and percolative transitions occurring at the same $T_c$, for
finite systems these effects have not yet merged. Indeed, the percolative contribution
appears as a second, much larger peak~\cite{RoOlAr15}, at a temperature 
significantly higher than $T_c$ (e.g., $T_2(L)\simeq 2T_c$ for $L=640$).
The height of this second peak behaves as $\Heq(T_2) \sim L^{d/\tau'(L)}$.
The exponent $\tau'$, associated with the height of the second
peak, is closer to the percolation value, $\taup=187/91\simeq 2.055$, but should crossover to $\taui$ as
the two peaks merge in the thermodynamic limit. 
The double peaked  heterogeneity is a property of
the geometric domains, while the physical (CK) domains, on the other hand, have
a single peak similar to the susceptibility.
Thus, for equilibrium finite samples, when describing the thermal and percolation
transitions with the cluster heterogeneity of geometric domains, they seem to be
disentangled, each one affecting the geometric properties more effectively at 
different temperatures. The main objective of this paper is to explore whether this new
measure may be useful, not only to study equilibrium properties of simple models, but
their dynamics as well.

After a quench from infinite to a below-critical temperature, the out-of-equilibrium dynamics
of the non conserved order parameter 2d Ising model is first attracted by the percolative critical
point and only then crosses over to the coarsening regime~\cite{ArBrCuSi07,SiArBrCu07}. In the
process, a percolation cluster first appears in the early stages ($t_{p_1}$) of the dynamics~\cite{ArBrCuSi07,SiArBrCu07},
but only becomes stable on a longer, size dependent timescale
$\tp\sim L^{\zp}$~\cite{BlCoCuPi14,BlCuPiTa17} where the exponent has been conjectured to be
$\zp=2/5$~\cite{BlCuPiTa17} for the square lattice.
This initially percolating state strongly correlates with the asymptotic state~\cite{DeOlSt96,Lipowski99,SpKrRe01a,SpKrRe01b,BaKrRe09,OlKrRe12,BlPi13,BlCoCuPi14}.
Besides eventually leading to the ground state, the (single flip) dynamics may halt at an on-axis striped configuration or slowly evolve
towards the ground state through a long lasting diagonally striped state. While the
timescale for the former case is $t_{\scriptscriptstyle \rm equil}/L^2\sim {\cal O}(1)$, the later is much longer,
$t_{\scriptscriptstyle \rm equil}/L^2\sim {\cal O}(L)$~\cite{Lipowski99}.
During the evolution, as the domains keep decreasing the excess energy at the curved interfaces, there
 appears a growing characteristic length associated with the coarsening regime,
 $\ell_d(t)\sim t^{1/z_d}$, with $z_d=2$~\cite{Bray94}.

The existence of a characteristic length obviously does not  imply that the system
is homogeneous, with domains similarly sized. A possible measure of the diversity
of the actual sizes is the cluster size heterogeneity previously discussed,
extended here to out of equilibrium configurations.  We are interested here
in its time evolution when applying different temperature protocols.
While both the initial and
the asymptotic equilibrium values of $\Heq$ have been measured~\cite{RoOlAr15},
there are many questions related to the intermediate time evolution of $H(t)$.
In particular, since $\Heq$ seems very responsive to the percolative equilibrium properties,
does the dynamical
size heterogeneity give information on the two regimes, approaching and departing
from the critical percolation point, before the dynamics is dominated by coarsening?
How distinct are these regimes?
Is $H(t)$ monotonic in time or is
there one or two peaks related to the equilibrium behavior?  Is it possible for a single parameter to give information on the two lengthscales associated with coarsening?  How different initial and final temperatures change the behavior? 
We address some of these questions, showing that the dynamical cluster size heterogeneity, $H(t)$, is indeed a suitable observable that not only distinguishes among different dynamical regimes, but also provides quantitative access to the scaling laws related to the growth of correlations and of percolative clusters during the dynamics. Furthermore, we show that the time evolution of $H(t)$ correlates with the nature of the correlations present in the initial state, whether long-range if the quench is performed from $T_0=T_c$ or absent, from $T_0 \to \infty$. It is also noteworthy that the short-time regime of $H(t)$ resulting from the dynamics triggered from $T_0 \to \infty$ to $T = 0$ allows a quantitative connection with the percolation-related peak observed in the equilibrium heterogeneity, $\Heq(T_2)$.



\section{Dynamical Cluster Heterogeneity}
\label{section.H}

Following different temperature quench protocols that drive the system out of
equilibrium, we study the 2d Ising model 
whose Hamiltonian is
\begin{equation}
{\cal H}= -J\sum_{<ij>} \sigma_i\sigma_j,
\end{equation}
where $J>0$, $\sigma_i=\pm 1$ is the spin at site $i$
and the sum is over all nearest neighbors sites on an $L\times L$ 
square lattice with periodic boundary conditions ($L$ is measured in
units of the lattice spacing $\ell_0$).
We choose the initial temperature $T_0$ to be either infinite or the
critical one, these equilibrium states thus differing by having  zero or infinite
range correlations, respectively. The fixed temperature adopted after the quench is $T=0$.
The simulations were performed on square lattices with linear sizes up to $L=5120$.
Averages up to 1000 samples were taken for the smaller systems while larger sizes 
require fewer samples (100). When $T_0=T_c$ it is necessary to equilibrate the 
system, and 1000 Swendsen-Wang steps were performed, while during the subsequent temporal
dynamics, in all cases, a fast version of the single-spin Glauber algorithm at $T=0$ was used~\cite{NeBa99}. 
Time is measured in Monte Carlo steps (MCS), where one unit corresponds
to $N$ attempts to flip.

\begin{figure}[htb]
\includegraphics[width=7cm]{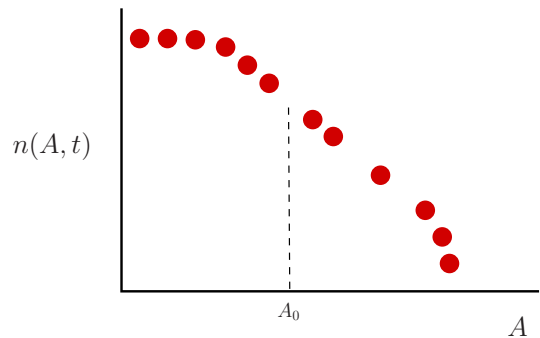}\
\caption{Schematic cluster size distribution for a single, finite sample. Differently from an infinite system or averaged distribution, some sizes have no realizations.
 The smallest missing size, $A_0$, is indicated by a vertical dashed line,
 separating the dense region of the distribution, $A<A_0$, from the sparse one, $A>A_0$. Those
 sizes that are indeed present in the specific configuration define a measure of the cluster
 size heterogeneity (in this example, $H=12$).}
\label{fig.defH}
\end{figure}

Along the time evolution of the system, we measure the dynamical cluster size
heterogeneity $H(t)$, taking into account only the non-spanning clusters (the
presence of one or more spanning clusters does not have a large influence on $H(t)$,
except close to the asymptotic state where it is small). It is defined, as in the
equilibrium case, as the number of different cluster sizes present at time $t$ on
a finite-size configuration (see the schematic depiction in Fig.~\ref{fig.defH}). 
Although different domain definitions are possible, we
here consider only geometrical domains, i.e., sets of connected parallel spins.


\subsection{Quench from $\Tc$ to $\Tf=0$}
\label{sec.TcT0}

\begin{figure*}[htb]
\includegraphics[width=8cm]{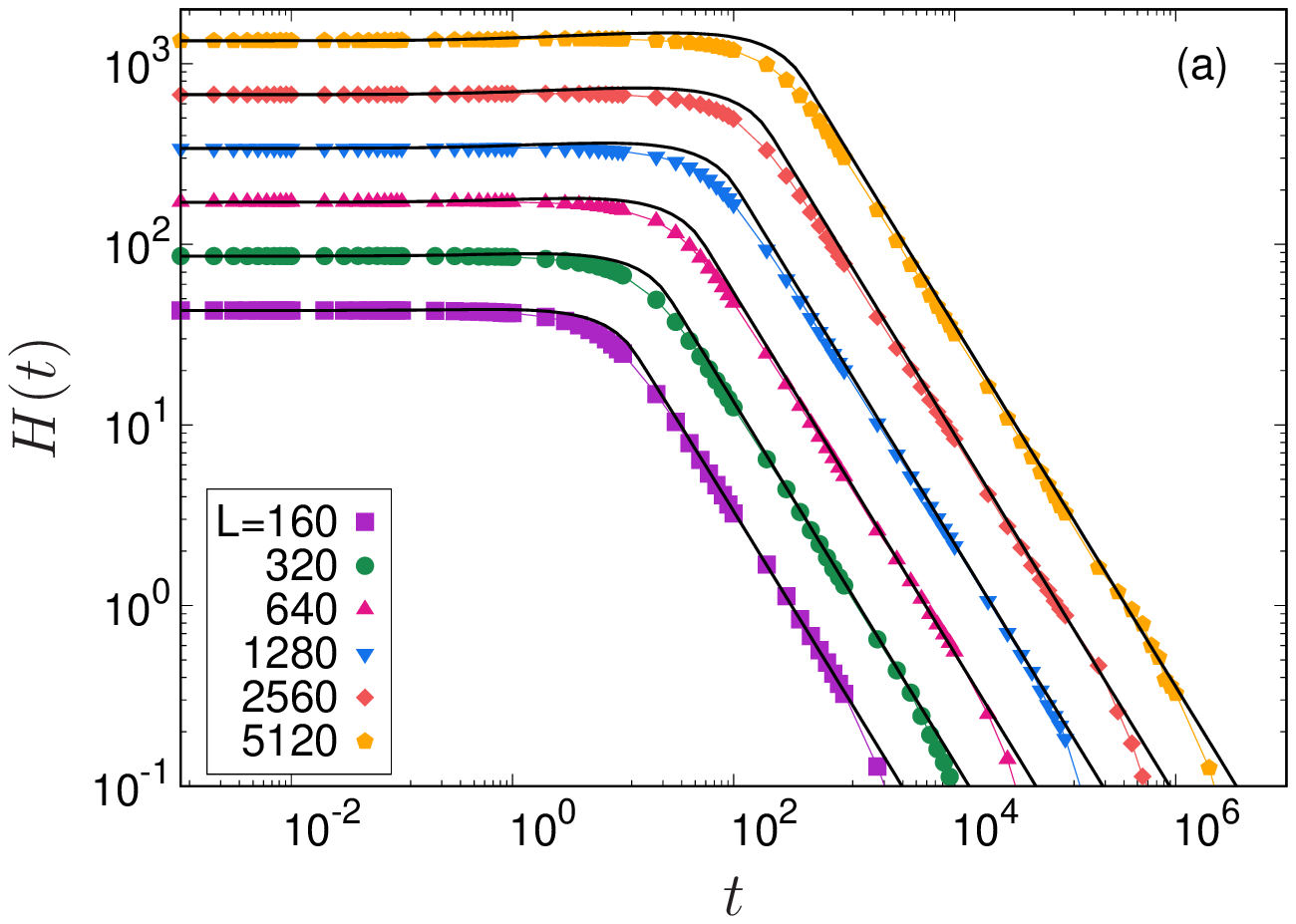}
\includegraphics[width=8cm]{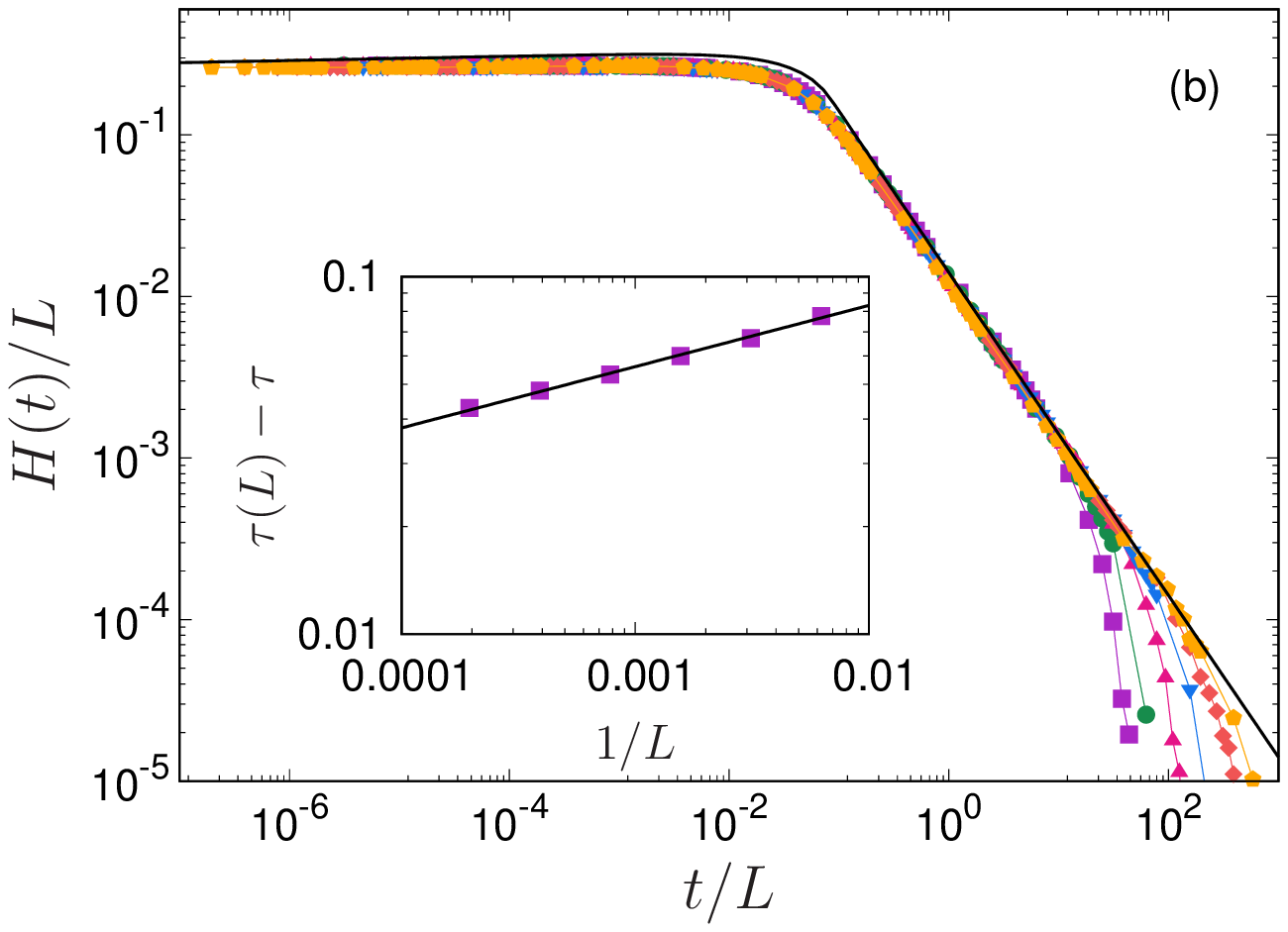}
\caption{Dynamical cluster size heterogeneity, $H(t)$, as a function of time (in MCS) 
after a temperature quench from $T_0=T_c$ down to $T_f=0$. For simplicity, only those samples that converged to a fully magnetized state were considered. a) In the first
  regime, $t\leq t_0$, $H(t)$ presents a slow variation starting from $\Heq(T_c)$, Eq.~(\ref{eq.HeqTc}), while in the power-law regime, the behavior is $t^{-1}$.
  The whole behavior is well approximated by Eq.~(\ref{eq.HtTc}) and shown as solid lines. We consider $c\simeq 0.029$, $\lambda\simeq 2$ and $\tau$, to take finite size effects into account,  as a fitting parameter.   b) The simulation data collapses well onto a universal curve, $H(t)=Lf(t/L)$,  using the asymptotic value of the Fisher exponent, $\taui=379/187$. The agreement
  with Eq.~(\ref{eq.HtTc}) is also very good. The inset shows that the values of $\tau(L)$ that
better fit the data in the panel (a) do converge to the correct value as $L$ increases.}
\label{fig.TcT0}
\end{figure*}

\begin{figure}[htb]
\includegraphics[width=8cm]{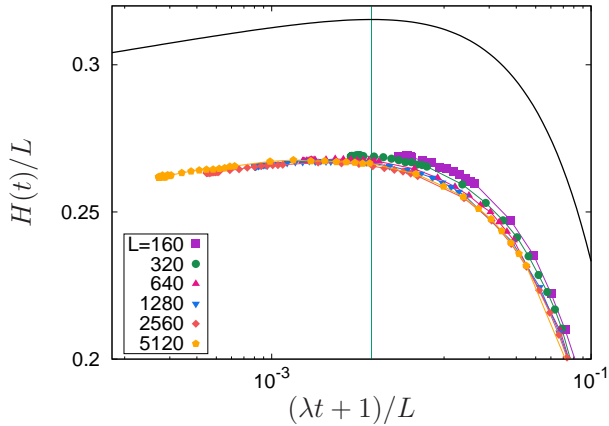}
\caption{Zoom into the region close to the end of the plateau, after a quench from $T_0=T_c$, showing a very small peak at $t_{\scriptscriptstyle\rm max}$ (vertical line), Eq.~(\ref{eq.Tc_peak}). The height of the peak in Eq.~(\ref{eq.HtTc}) depends on the precise value of $c$ and the difference to the numerical data appears larger because of the chosen scale.}
\label{fig.TcT0_peak}
\end{figure}

After a quench from the equilibrium initial state at the
Ising critical temperature ($T_0=\Tc$), the cluster size distribution of geometric domains
evolves as~\cite{ArBrCuSi07,SiArBrCu07}
\begin{equation}
n(A,t) \simeq \frac{c[ \lambda(t+t')]^{\taui-2}}{[A+\lambda(t+t')]^{\taui}},
\label{eq.nA}
\end{equation}
where $n(A,t)dA$ is the average number of (non spanning) clusters,  per unit area, whose size is
between $A$ and $A+dA$.  The constant
$c\simeq 0.029$~\cite{ArBrCuSi07,SiArBrCu07} is very close to the
Cardy-Ziff number~\cite{CaZi03}, $c_h=1/(8\pi\sqrt{3})\simeq 0.023$,  $\lambda\simeq 2$ (time and area units are unitary) is a
temperature dependent material constant (the chosen value is for $T=0$) and $t'$ is a microscopic time such that $\lambda t'\simeq 1$.
The above distribution is an average while the heterogeneity $H$ is measured from single configurations as schematically
shown in Fig.~\ref{fig.defH}. 
At the moment of the quench, there is an already stable spanning cluster,
$t_p\simeq t_{p_1}\simeq 0$~\cite{BlCoCuPi14}. 
Differently from the averaged distribution of Eq.~(\ref{eq.nA}), for a single sample there are holes in the distribution, as not all possible
sizes may be present. Denoting by $A_0(t)$ the smallest missing size at time $t$, the sample cluster size distribution is dense for $A<A_0(t)$ and sparse above it. If $\ell_0$ is a microscopic length and $(\ell_0L)^2$ is the area of the system, $A_0$ is the size such that $(L\ell_0)^2n(A_0,t)\ell_0^2\sim 1$. Thus, setting $\ell_0=1$, we obtain that
\begin{equation}
A_0(t) \simeq(\lambda t+1) \left[\left(\frac{L \sqrt{c}}{\lambda t+1}\right)^{2/\tau}-1\right]\Theta (t-t_0)
\label{eq.A0}
\end{equation}
and the dense region of the cluster size distribution, on average, disappears after
a time 
\begin{equation}
t_0\simeq \frac{L\sqrt{c}}{\lambda}.
\end{equation}
The cluster size heterogeneity after the quench,
$H(t)$, is shown in Fig.~\ref{fig.TcT0}a for different lattice sizes.
The coarsening process moves the whole distribution to the left, removing the smallest
clusters, initially changing very slowly the value of $H(t)$ up to $t\simeq t_0$.
It then crosses over to a different regime, decreasing as a power
law, when the dense region is about to disappear. Once the remaining distribution is sparse,
almost all present cluster sizes appear only once, and $H(t)$ becomes equivalent to the number
of clusters. These two contributions to $H(t)$ may be approximated by
\begin{equation}
H(t) \simeq A_0 + L^2\int_{A_0}^{\infty} dA\; n(A,t),
\label{eq.Htanal}
\end{equation}
where the first and second terms correspond, respectively, to the size of the dense region
and the number of clusters in the sparse one. Using Eqs.~(\ref{eq.nA}) and (\ref{eq.A0})
with Eq.~(\ref{eq.Htanal}), we get an expression for $H(t)$ at all times:
\begin{equation}
  H(t) \simeq \left\{\!\!
  \begin{array}{l}
    \displaystyle(\lambda t+1) \left[ \frac{\tau}{\tau-1}\left(\frac{L \sqrt{c}}{\lambda t+1}\right)^{2/\tau}\!\!-1\right]  ,\; t\leq t_0 \\
 \\
    \displaystyle\frac{L^2 c}{\tau-1}\frac{1}{\lambda t+1}        ,\; t\geq t_0.
  \end{array}
  \right.
\label{eq.HtTc}
\end{equation}
Notice that $H(t\to\infty)=0$ in the above expression because, in our definition, the spanning clusters are not accounted for. At $t=t_0\simeq L\sqrt{c}/\lambda$,
both terms give $H(t_0)\simeq L\sqrt{c}/(\tau-1)$ while the initial value,
corresponding to the equilibrium state at $T_c$, is
\begin{equation}
  H(0)\simeq \Heq(T_c)\simeq \frac{\tau}{\tau-1} c^{1/\tau}L^{2/\tau}.
\label{eq.HeqTc}
\end{equation}
Fig.~\ref{fig.TcT0}a also compares the simulations with the above expression for $H(t)$ as solid lines. The agreement is pretty good, except where there is a change of regime, close to $t_0$, where $n(A,t)$ is still significant and there may be more than one cluster with the same size, 
originating the small deviation seen in Figs.~\ref{fig.TcT0}a and \ref{fig.TcT0_peak}.  Despite its exact value being known, we have
considered $\tau$ as a fitting parameter in order to take finite size effects into account. The inset of Fig.~\ref{fig.TcT0}b shows the values of $\tau(L)$ obtained from each fit (performed only for the initial times, $t<10^{-2}$) and how they converge to $\taui=379/187$ as $L\to\infty$. Fig.~\ref{fig.TcT0}b also shows
that the same data, when properly rescaled, present an excellent collapse. From Eq.~(\ref{eq.HtTc}) we see that the rescaling is
$H(t)=L^{-1}f(t/L)$, where $f(x)\sim x^{-1}$, a power-law decay, for $x\gg 1$ and
$x^{1-2/\tau}$, a very slow increase, for $x\ll 1$. There is a further, subtle feature of the
numerical data, again well captured by Eq.~(\ref{eq.HtTc}), that can be seen in
Fig.~\ref{fig.TcT0_peak}: $H(t)$ is not a monotonous function. It presents a maximum
$H(t_{\scriptscriptstyle\rm max})$
whose location agrees well with the numerical data,
\begin{equation}
  \frac{\lambda t_{\scriptscriptstyle\rm max}+1}{L} = \sqrt{c}\left(\frac{\tau-2}{\tau-1}\right)^{\tau/2}\simeq 0.004,
\label{eq.Tc_peak}
\end{equation}
although the height has a small deviation (enlarged in Fig.~\ref{fig.TcT0_peak} because of
the chosen vertical scale).

\subsection{Quench from $T_0\to\infty$ to $\Tf=0$}
\label{sec.TinfT0}

\begin{figure*}[htb]
\includegraphics[width=8cm]{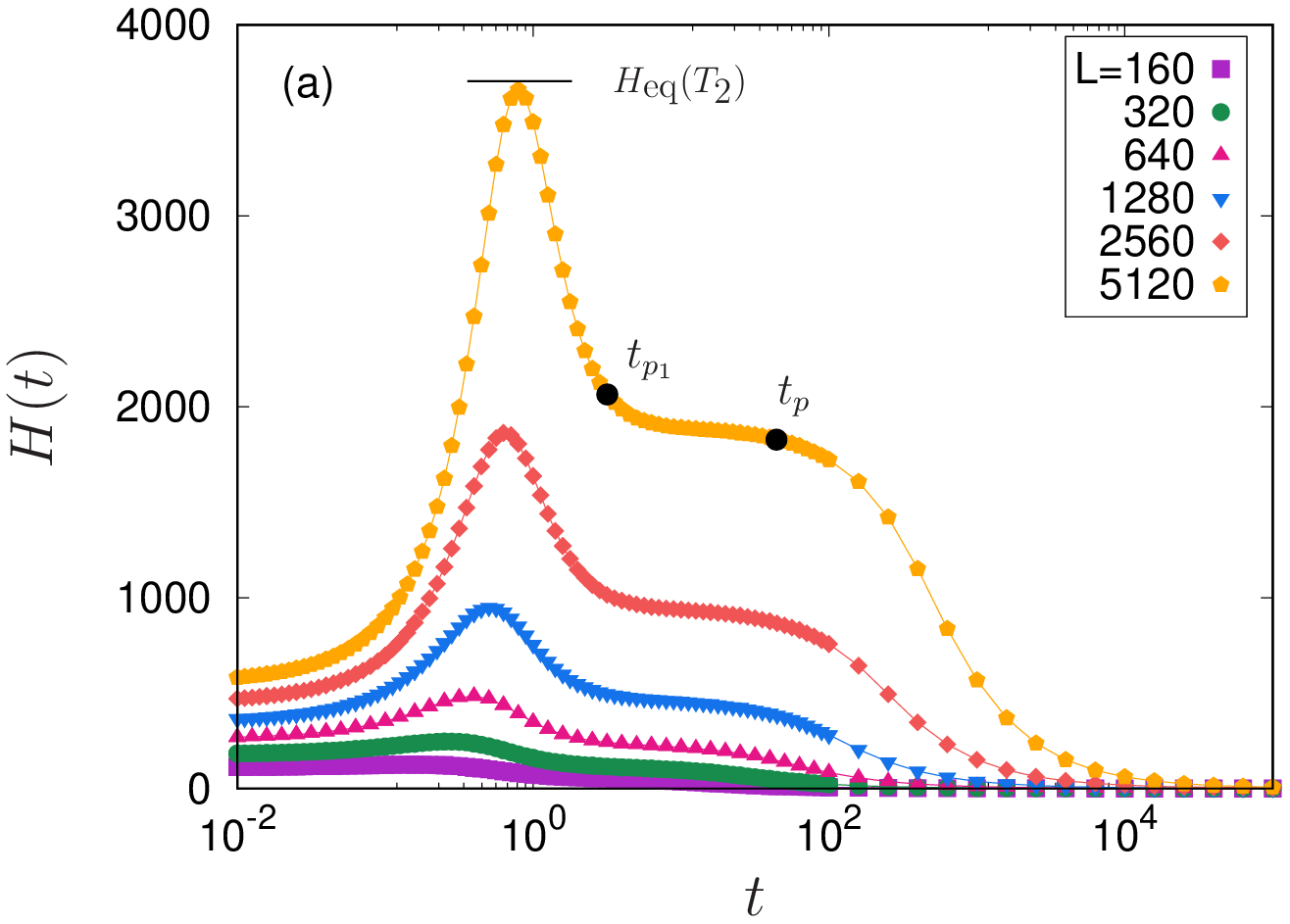}
\includegraphics[width=8cm]{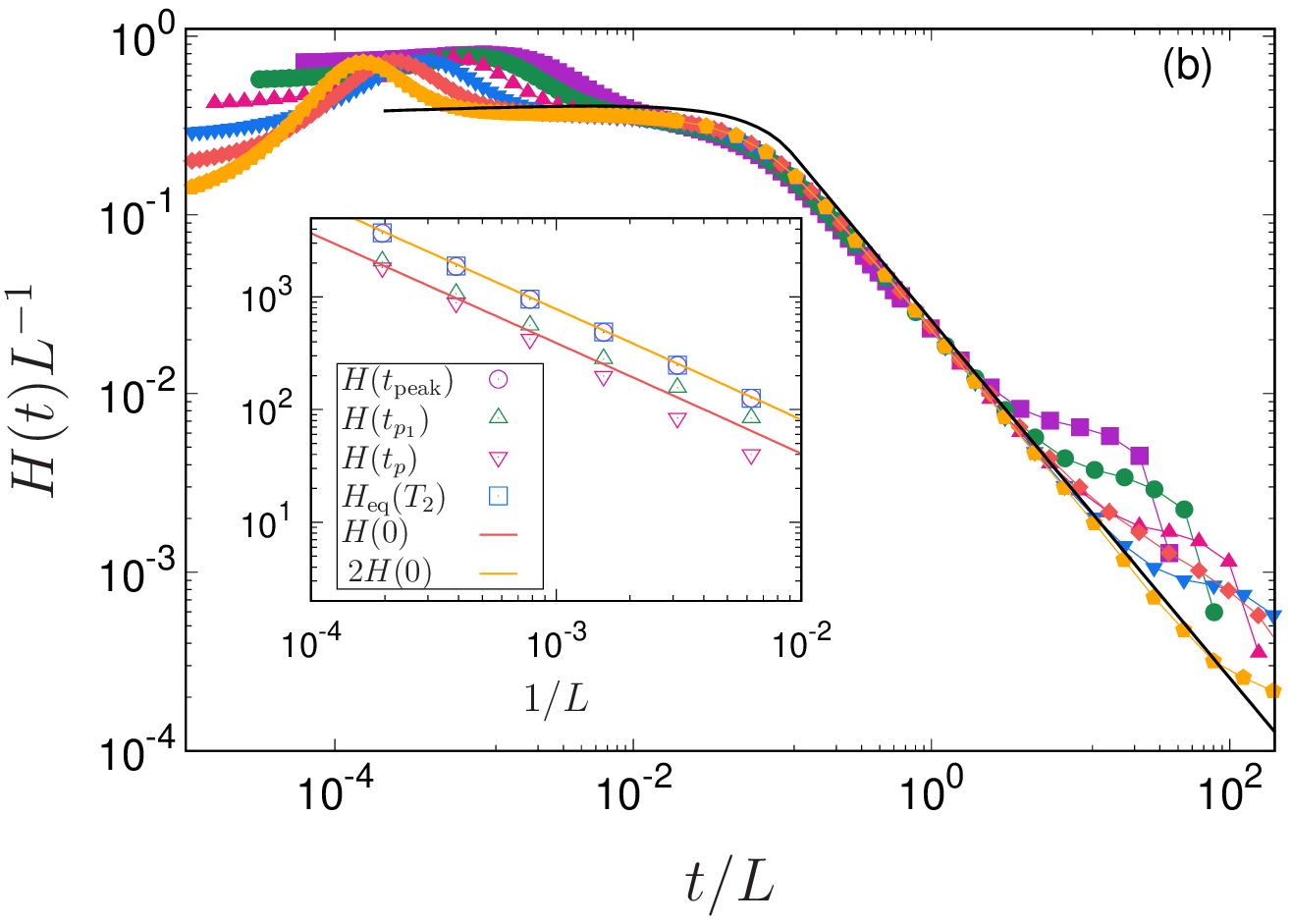}
\caption{a) Dynamical cluster size heterogeneity $H(t)$ as a function of time (in MCS) after a temperature quench from $T_0\to\infty$ down to $T_f=0$.  For simplicity, only those samples that converged to a fully magnetized state,were considered. For the largest size, we indicate the times when a percolating cluster first appears and when it becomes stable, $t_{p_1}$ and $t_p$, respectively. In Ref.~\cite{RoOlAr15} it was observed that the equilibrium heterogeneity, $\Heq$, has a second, larger peak at $T_2(L)$, well above $T_1\simeq T_c$ where another, smaller peak is located. The value of $\Heq(T_2)$ well agrees with $H(\tpeak)$ and is shown, for the largest simulated size only, 
as a small horizontal line on top of the peak. b) Data collapse. As the system size increases, a region  between $t_{p_1}$ and
$t_p$, where $H(t)$ slowly changes, becomes apparent. The behavior for $t>t_{p_1}$ is well approximated by Eq.~(\ref{eq.HtTinf}) with $\taup=187/91$ (solid line). The inset shows, in the upper straight line, the height of the peaks $H(\tpeak)$ and $\Heq(T_2)$, indistinguishable at this scale, along with Eq.~(\ref{eq.Hpeak}), as a function of the system size. The data below (triangles) correspond to the values of $H(t_p)$ and $H(t_{p_1})$ that, albeit different, get close as $L$ increases.}
\label{fig.TinfT0}
\end{figure*}

When quenching the system from $T_0\to\infty$ down to
$\Tf=0$, the general cluster size heterogeneity behavior can be seen in Fig.~\ref{fig.TinfT0}a. 
At the initial, high temperature, despite spins being uncorrelated, small clusters of parallel spins
are present. The initial heterogeneity is not very large and slowly grows with
the system size, $\Heq(T\to\infty)\sim \ln L$~\cite{NoLePa11,JoYiBaKi12,RoOlAr15}, as
can be observed in Fig.~\ref{fig.TinfT0}a.
Soon after the quench, $H(t)$ presents
a pronounced peak followed by a growing, intermediate plateau before the final power-law
decrease toward the asymptotic state. The dynamics is eventually attracted~\cite{DeOlSt96,SpKrRe01a,SpKrRe01b,OlKrRe12}
to a state that is either fully magnetized or contains on or off-axis stripes.
Although the results are similar, for simplicity we kept only those samples that got,
eventually, fully magnetized.

Differently from the previous case, the initial, equilibrium state at $T_0\to\infty$ is not
critical. Nonetheless, before entering the scaling regime, the system first approaches the random site 
percolation critical state~\cite{ArBrCuSi07,SiArBrCu07}, with an average cluster size distribution given by a power-law
$A^{-\taup}$ whose Fisher exponent is $\taup=187/91$. As discussed in the introduction, the first occurrence of a percolating, albeit
 unstable, cluster is at the early time $t_{p_1}$, while at $t_p$ it becomes
stable. After the cluster size distribution becomes critical at $t_{p_1}$,
its time evolution is well approximated by~\cite{ArBrCuSi07,SiArBrCu07,BlCuPiTa17}
\begin{equation}
n(A,t) \simeq \frac{2c[ \lambda(t+t_{p_1}+t')]^{\taup-2}}{[A+\lambda(t+t_{p_1}+t')]^{\taup}},
\label{eq.nAp}
\end{equation}
where the factor 2 in the numerator comes from the existence of clusters with both
positive and negative magnetizations while in the related percolation problem, only
particle clusters, not voids, are accounted for. In analogy to the previous case, the behavior of $H(t)$ after
a quench from $T_0\to\infty$, calculated using Eq.~(\ref{eq.Htanal}), is given by
\begin{equation}
  H(t) \!\simeq\! \left\{
  \begin{array}{l}
    \!\displaystyle(\lambda t+1) \left[ \frac{\taup}{\taup-1}\!\left(\frac{L \sqrt{2c}}{\lambda t+1}\right)^{2/\taup}\!-\! 1\right],\; 
t_{p_1}\!<t\leq t_0 \\
 \\ \!
    \displaystyle\frac{2L^2 c}{\taup-1}\frac{1}{\lambda t+1} ,\; t\geq t_0,
  \end{array}
  \right.
\label{eq.HtTinf}
\end{equation}
where, in this case, $t_0\simeq L\sqrt{2c}/\lambda$. As in the $T_0=T_c$ case, $H(t)$ also
has a broad and small maximum before $t_0$, more precisely at $(\lambda t_{\scriptscriptstyle\rm max}+1)/L =\sqrt{2c}[(\taup-2)/(\taup-1)]^{\taup/2}\simeq 0.011$. However, this maximum does not appear in the simulation and $H(t)$ seems to always decrease. This is because, for the sizes considered here, the region $t\leq t_0$ may not be yet fully developed or, another possibility, because of the presence of the initial,
precursor maximum that reverses the behavior below $t_{p_1}$.

Since Eq.~(\ref{eq.nAp}) considers
an effective initial state at the percolation threshold, the above expression for
$H(t)$ is not expected to capture any feature before $t_{p_1}$. Indeed, $t=0$
corresponds to the beginning of the slowly changing region, that roughly extends between $t_{p_1}$ and $t_p$ (and
whose width depends on $L$), observed in Figs.~\ref{fig.TinfT0}a
and b:
\begin{equation}
  H(0)\simeq \frac{\taup}{\taup-1} (L\sqrt{2c})^{2/\taup}.
\end{equation}

In contrast with the $T_0=T_c$ case, $H(t)$ has a very pronounced peak just before the 
appearance of the first percolating cluster, i.e., $\tpeak<t_{p_1}$, that being a precursor
feature of the percolating state. After the quench,
as the correlations build larger clusters, the size distribution widens and $H(t)$ increases. However, as the largest cluster increases, less space remains for the other clusters. Thus, the maximum heterogeneity occurs slightly before $t_{p_1}$. Interestingly, the
height of the peak seems to correspond to the equilibrium heterogeneity at the
second peak observed in Ref.~\cite{RoOlAr15}, i.e., for each system size, 
$H(\tpeak)\simeq  \Heq(T_2)$, as
indicated by a small horizontal line in Fig.~\ref{fig.TinfT0}a (only for the largest $L$). Moreover,
we numerically observe that it is twice the height at $t_{p_1}$:
\begin{equation}
 H(\tpeak)\simeq  \Heq(T_2) \simeq 2H(0)\simeq \frac{2\taup}{\taup-1} (L\sqrt{2c})^{2/\taup}.
\label{eq.Hpeak}
\end{equation}
For $t\geq t_0$, the power-law behavior of $H(t)$ is similar to the $T_0=T_c$ case and
accounts for the number of clusters, differing only by the value of $\taup$ and the factor
 2 in the numerator.  The data for $t>t_{p_1}$ is well described by
 Eq.~(\ref{eq.HtTinf}), as can be seen in Fig.~\ref{fig.TinfT0}b.
 Nonetheless, by rescaling both $H(t)$ and time by $L$, Fig.~\ref{fig.TinfT0}b,
although the finite size effects are somewhat stronger than in the $T_0=T_c$ case,
both the agreement with Eq.~(\ref{eq.HtTinf}) and  the collapse in the same region
are very good. 

\begin{figure}[htb]
\includegraphics[width=8cm]{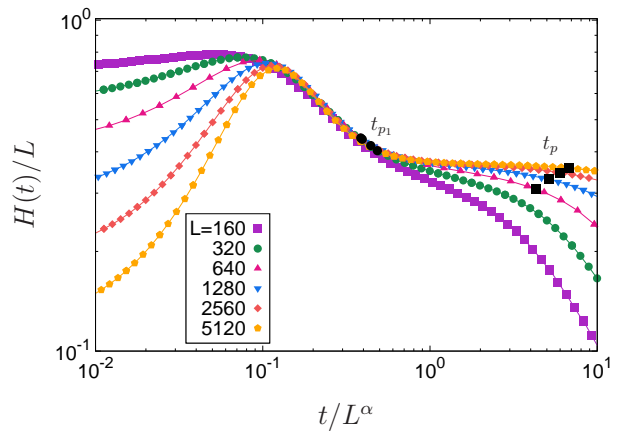}
\caption{Rescaling of the early time region near the peak of $H(t)$ after a temperature quench from $T_0\to\infty$. The region that includes both the peak and $t_{p_1}$ is well collapsed using
$\alpha\simeq 0.22$.}
\label{fig.TinfT0_peak}
\end{figure}

The strong, early peak is a precursor effect of the percolating cluster that appears
soon afterwards, $\tpeak<t_{p_1}$. Analogously, as the temperature is slowly
decreased, the equilibrium peak at $T_2$~\cite{RoOlAr15} also
anticipates the first appearance of a percolating cluster.
Interestingly, the data collapse in Fig.~\ref{fig.TinfT0}b fails in the very early
regime, indicating that the dynamical scaling length $\xi(t)\sim t^{1/2}$ is not the
sole relevant length scale after the quench. The precursor peak shifts to the left, indicating that a scaling
factor $L^{\alpha}$, with $\alpha<1$, should be considered instead of $L$. Indeed,
as seen in Fig.~\ref{fig.TinfT0_peak}, a good collapse around the peak is obtained with $\alpha\simeq 0.22$. However, notice that although  
the peaks are well collapsed, neither the black circles indicating $t_{p_1}$  nor
the black squares for $t_p$ present a good convergence. Different values of the
exponent $\alpha$ can, on the other hand, collapse those characteristic times. For
$t_p$, it was shown in Ref.~\cite{BlCuPiTa17} that the exponent is 0.4.


\begin{figure}[htb]
\includegraphics[width=8cm]{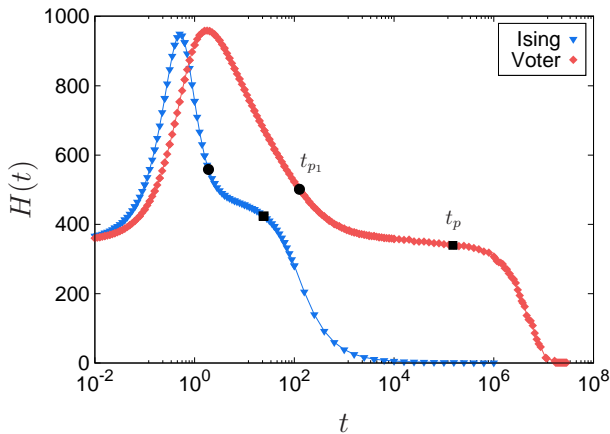}
\caption{Comparison between the time evolution of $H(t)$ for the Ising at $T=0$ and
  the Voter model, both starting from an initially uncorrelated state ($T_0\to\infty$). The
  system linear size is $L=1280$ and time is in MCS. Notice that the precursor peak has roughly
the same height for the two models, suggesting a common mechanism.}
\label{fig.voter}
\end{figure}

In order to check how universal the $H(t)$ behavior is, we compare in Fig.~\ref{fig.voter},
for $L=1280$, the behavior of $H(t)$ for the Ising model after a quench from $T_0\to\infty$ and
the Voter model (VM) evolving from a fully uncorrelated state. The VM is interesting as there is
no bulk noise and detailed balance is not obbeyed. Instead of considering the energy variation
for a putative flip, as in the Ising model, in the VM the spin choses and aligns with a single
neighbor. As shown in Ref.~\cite{TaCuPi18}, the timescales are all
larger in the VM, nonetheless, the overall behavior of $H(t)$ is similar, Fig.~\ref{fig.voter}.
Moreover,
defining $t_{p_1}$ and $t_p$ as above (even if the critical properties of the percolating
cluster do not correspond to critical percolation~\cite{TaCuPi18}, we can see in
Fig.~\ref{fig.voter} that they are related, respectivelly, to the end of the precursor peak
and the end of the plateau. A remarkable feature in this figure is the height of the early
peak, that is roughly the same in both models, suggesting a more general mechanism.


\section{Conclusions}

In equilibrium at high temperatures, domains of parallel neighboring
spins are not large and within a limited range of sizes, thus the number of different
domain sizes in a given configuration, $\Heq(T)$, is small. Decreasing the temperature,
spins become more correlated and the clusters
increase and diversify, increasing $\Heq$. However, as some clusters get comparable
to the size of the system, the lack of space tends to decrease the diversity. In the presence
of these competing mechanisms, one expects a peak in $\Heq$. Remarkably, for
geometric domains, two peaks are
present~\cite{RoOlAr15}, one near the temperature where the percolating cluster first appears~\cite{RiCuPi18}
and a second one very close to $\Tc$. 
We here extended this equilibrium measure of how heterogeneous the
domains are in size, to non-equilibrium situations, $H(t)$. Specifically, we explore its usefulness in the non conserved order parameter dynamics of the 2d Ising model after a
sudden quench in temperature, confirming that this observable unveils the rich interplay between percolation and ferromagnetism either close to the phase transition (equilibrium) or at short-time scales during the dynamics.

For quenches starting at $T_c$, $H(t)$ presents an initial plateau that increases very
slowly, attaining a shallow maximum before crossing over to a power-law behavior.
In this latter regime, the sample size distribution is very sparse and the probability
of two domains having the same size is small. The heterogeneity does correspond, in this
time regime, to the total number
of clusters and decays as a power law. When the system, instead, is first equilibrated
at $T_0\to\infty$ (random spin configuration), in addition to these regimes {\it after} it passes through the
percolation critical point, there is also a very pronounced peak that is a precursor
indication that a giant, percolating cluster is being built. 


The rich behavior of $H(t)$ suggests that it would be interesting to consider several extensions,
both in equilibrium and after a quench in temperature.
While for the Ising model each domain has a single neighbor and its size decreases at
the same, constant rate, for the ($q>2$) Potts model domains may either decrease or increase as their time evolution, given by the von Neumann law, depends on their number of sides. The
coarsening behavior is thus richer~\cite{LoArCuSi10,LoArCu12,DeRe19}.
As a consequence, domains with the same area but different number of sides have a larger
probability of evolving into different sizes, increasing the heterogeneity. Such
mechanism, that breaks the degenerescency of areas depending on the number of sides
is absent in the Ising model. Another interesting cases are the Ising model with conserved
order parameter~\cite{SiSaArBrCu09,TaMiDe15,TaCuPi18} or disorder~\cite{SiArBrCu08,CoCuInPi17,CoCuInPi19}. Although we focused here on geometric domains,
the heterogeneity associated with the Coniglio-Klein clusters~\cite{CoKl80} would also be of interest~\cite{RoOlAr15}, along with the heterogeneity of perimeters. 
The dynamics of the 3d Ising model is more challenging~\cite{SpKrRe01a,ArCuPi15,VaChDa19}, as
multiple frozen percolating clusters coexist and, for sufficiently large systems, the
ground state is never reached. In addition, the thermal and percolation transitions do not coincide. Finally, it would be important to verify our results in experimental setups~\cite{SiArDiBrCuMaAlPi08,Almeida20}.
 



\begin{acknowledgments}
We are grateful to Renan Almeida, Federico Corberi, Leticia Cugliandolo, Marco Picco and Hiromitsu Takeuchi for useful conversations and/or
comments on the manuscript.
Work partially supported by the Brazilian agencies FAPERGS, FAPERJ, CNPq, and CAPES. 
\end{acknowledgments}


\end{document}